\documentclass[iop,apjl]{emulateapj}
\usepackage{amsmath}
\usepackage{amssymb}
\usepackage{verbatim}
\usepackage{natbib}
\usepackage{txfonts}
\usepackage{graphicx}
\usepackage{url}
\usepackage{multirow}
\usepackage{subfigure}

\newcommand{\til}{\raisebox{-0.7ex}{\ensuremath{\tilde{\;}}}}
\newcommand{\as}{$^{\prime\prime}$}
\newcommand{\am}{$^{\prime}$}
\newcommand{\dg}{$^{\circ}$}
\newcommand{\h}{$^{\mathrm h}$}
\newcommand{\m}{$^{\mathrm m}$}
\newcommand{\s}{$^{\mathrm s}$}

\begin{document}

\title{A Highly Eccentric 3.9-Millisecond Binary Pulsar in the Globular Cluster NGC\,6652}
\author{Megan E. DeCesar}
\affil{Physics Department, University of Wisconsin-Milwaukee, 1900 East Kenwood Boulevard, Milwaukee, WI 53211, USA}
\author{Scott M. Ransom}
\affil{National Radio Astronomy Observatory, Charlottesville, VA 22903, USA}
\author{David L. Kaplan}
\affil{Physics Department, University of Wisconsin-Milwaukee, 1900 East Kenwood Boulevard, Milwaukee, WI 53211, USA}
\author{Paul S. Ray}
\affil{Space Science Division, Naval Research Laboratory, Washington, DC 20375-5352, USA}
\author{Aaron M. Geller}
\affil{Center for Interdisciplinary Exploration and Research in
Astrophysics (CIERA) and Department of Physics and Astronomy,
Northwestern University, 2145 Sheridan Road, Evanston, IL 60208, USA}
\affil{Department of Astronomy and Astrophysics, University of
Chicago, 5640 S. Ellis Avenue, Chicago, IL 60637, USA}

\begin{abstract}
We present the Robert C. Byrd Green Bank Telescope discovery of the
highly eccentric binary millisecond pulsar PSR\,J1835$-$3259A in the
\textit{Fermi} Large Area Telescope-detected globular cluster
NGC\,6652.  Timing over one orbit yields the pulse period 3.89\,ms,
orbital period 9.25\,d, eccentricity $\sim$\,$0.95$, and an unusually
high companion mass of 0.74\,$M_{\odot}$ assuming a 1.4\,$M_{\odot}$
pulsar.  We caution that the lack of data near periastron prevents a
precise measurement of the eccentricity, and that further timing is
necessary to constrain this and the other orbital parameters. From
tidal considerations, we find that the companion must be a compact
object. This system likely formed through an exchange encounter in the
dense cluster environment.  Our initial timing results predict the
measurements of at least two post-Keplerian parameters with long-term
phase-connected timing: the rate of periastron advance
$\dot{\omega}$\,$\sim$\,$0.1\degr$\,yr$^{-1}$, requiring 1\,yr of
phase connection; and the Einstein delay
$\gamma_{\mathrm{GR}}$\,$\sim$\,10\,ms, requiring 2--3\,yr of timing.
For an orbital inclination $i > 50^{\circ}$, a measurement of
$\sin{i}$ is also likely. PSR\,J1835$-$3259A thus provides an
opportunity to measure the neutron star mass with high precision; to
probe the cluster environment; and, depending on the nature of the
companion, to investigate the limits of general relativity.
\end{abstract}

\section{Introduction \label{sec:intro}}

Globular clusters (GCs) are efficient producers of low-mass X-ray
binaries (LMXBs) and their descendant millisecond pulsars
\citep[MSPs;][and references therein]{Papitto+2013}: orders of
magnitude more MSPs and LMXBs exist, by mass, in clusters than in the
Galactic field \citep{CamiloRasio2005}.  The dense GC environment
heightens the probability of stellar interactions \citep[parameterized
  by encounter rate $\gamma$;][]{VerbuntFreire2014},
increasing the likelihood of forming new binaries and of existing
binaries gaining new companions.  Systems that rarely (if ever) form
through known binary evolutionary channels in the field can in
principle form through such stellar interactions in GCs, for example:
sub-ms pulsars; highly eccentric binaries; or unusual binary systems
like MSP-main sequence \citep[][and references
  therein]{Pallanca+2010}, MSP-MSP, or MSP-black hole (MSP-BH)
binaries \citep{Ransom2008a}.  Such systems would allow astrophysical
studies that may not otherwise be possible, e.g., strong-field tests
of gravity with MSP-MSP or MSP-BH binaries.

The \textit{Fermi} Large Area Telescope (LAT) has found MSPs to be
nearly ubiquitous $\gamma$-ray emitters;
GeV emission from GCs
\citep{Abdo+2010, Tam+2011}
may originate from the clusters' MSP populations
\citep[e.g.,][]{VenterDJC2009}. The LAT-detected GCs included
NGC\,6388 and NGC\,6652 \citep{Abdo+2010}, neither of which contained
any known MSPs, but whose $\gamma$-ray luminosities implied large MSP
populations.  NGC\,6388 is particularly interesting due to its high
encounter rate \citep[e.g.,][]{Freire+2008b, Maxwell+2012}; NGC\,6652
may also have a higher encounter rate than previously thought
\citep{NoyolaGebhardt2006}. The presence of a MSP population is
supported by the number of X-ray sources, including LMXBs, in both
clusters \citep[at least two in NGC\,6652 and five in
  NGC\,6388:][]{Maxwell+2012, Stacey+2012}.  We searched these
clusters for radio pulsars; here we report on the discovery and timing
of a highly eccentric binary MSP in NGC\,6652.

\section{Observations and Pulsar Search Analysis}
We observed NGC\,6388 and NGC\,6652 (Table~\ref{tab:targets}) with the
National Radio Astronomy Observatory\footnote{The National Radio
  Astronomy Observatory is a facility of the National Science
  Foundation operated under cooperative agreement by Associated
  Universities, Inc.} Robert C. Byrd Green Bank Telescope (GBT) and
the Green Bank Ultimate Pulsar Processing Instrument
(GUPPI\footnote{\url{http://dx.doi.org/10.1117/12.790003}}) backend
\citep{DuPlain+2008}, at S band (2\,GHz) in coherent search mode.  The
data were taken with 2048 spectroscopic channels and an effective
bandwidth of 700\,MHz (accounting for excised radio frequency
interference), with coherent dedispersion at the clusters'
predicted dispersion measure (DM) values (Table~\ref{tab:targets}).
The observing log and minimum detectable flux densities
$S_{\mathrm{min}}$ for an assumed 10\% duty cycle are given
in Table~\ref{tab:observations}.

The data were analyzed using
\textsc{presto}\footnote{\url{http://www.cv.nrao.edu/~sransom/presto/}}
\citep{Ransom2001}.  Time series were dedispersed at 1900 DMs from
0--691\,cm$^{-3}$\,pc for NGC\,6652 and 5456 DMs between
0--800\,cm$^{-3}$\,pc for NGC\,6388, and were searched for
periodicities.  We searched for accelerated signals over $z = \pm200$
Fourier bins \citep[cf.][]{Ransom+2001}, corresponding to maximum
line-of-sight accelerations between
$\pm$\,300--3000\,cm\,s$^{-2}$ for a 5\,ms pulsar.

\section{Discovery and Initial Timing Analysis of PSR J1835$-$3259A \label{sec:discovery}}
\begin{figure}
\begin{center}
\includegraphics[scale=0.5, angle=-90]{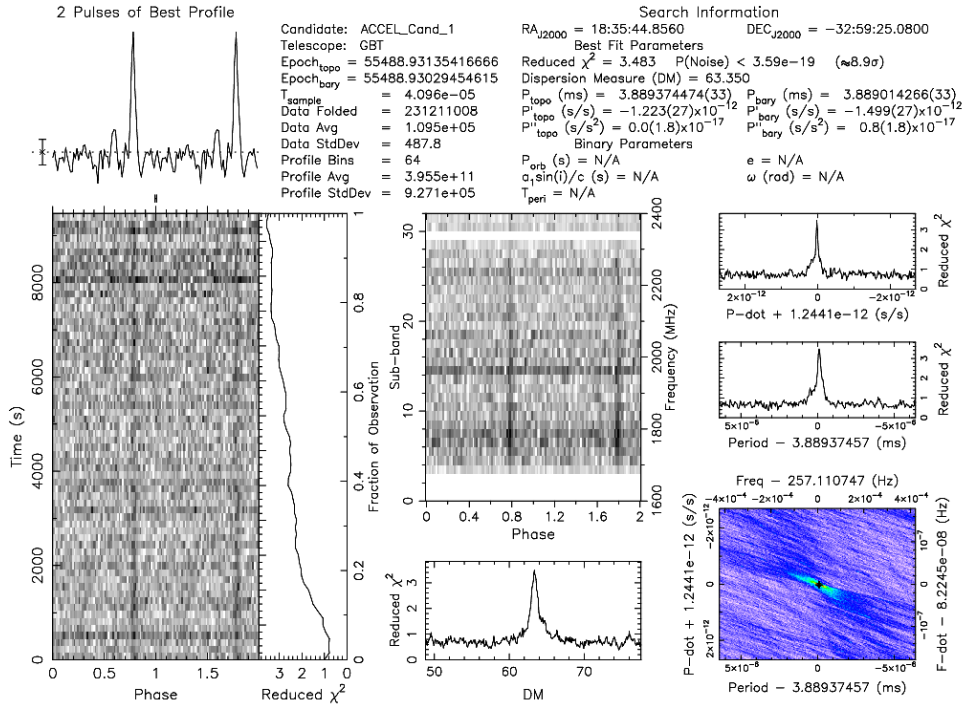}
\caption{\textsc{presto} discovery plot. \textit{Top left:} Two cycles of the
  summed pulse profile.  \textit{Left:} Two pulse cycles, with the
  signal in 64 sub-integrations split evenly over the full integration
  time. The $\chi^2$ plot on the right shows the steadily increasing
  S/N over time. \textit{Middle:} Two pulse cycles with the signal in
  32 frequency sub-bands split evenly over the full
  bandwidth. \textit{Bottom middle:} The $\chi^2$ increases
  dramatically near the pulsar DM. \textit{Right:} The maximum
  $\chi^2$ determines the best $\dot{P}$ \textit{(top)} and $P$
  \textit{(middle)}. \textit{Bottom right:} Covariance between $P$ and
  $\dot{P}$.}
\label{fig:discovery}
\end{center}
\end{figure}

We discovered PSR\,J1835$-$3259A (hereafter NGC6652A) in the direction
of NGC\,6652 \citep{DeCesarRR2011}, with the fundamental frequency
at an acceleration of 11.1\,cm\,s$^{-2}$ ($z = 9$).
Figure~\ref{fig:discovery} shows the \textsc{presto} discovery plot,
and Table~\ref{tab:observations} contains estimates of the 2\,GHz flux
density $S_2$.
We discuss the unexpectedly low DM value of 63.35\,cm$^{-3}$\,pc below
(Section~\ref{sec:membership}).

We fit the Doppler-shifted $P$ and $\dot{P}$
(Table~\ref{tab:observations}) with a phase-incoherent orbital model
 \citet{FreireKL2001}, using a routine by R. Lynch (private
communication) employing
\textsc{mpfit}\footnote{\url{http://cars9.uchicago.edu/software/python/mpfit.html}},
and found a very eccentric orbit ($e > 0.7$).  Starting with this
orbital model, we ran
\textsc{tempo}\footnote{\url{http://tempo.sourceforge.net/}}
iteratively on the pulse times of arrival (TOAs;
Table~\ref{tab:observations}) to converge on a family of timing
solutions.  We phase-connected the first five observations; we did not
observe the pulsar at periastron, between observations 5 and 6, so we
allowed the phase between these observations to remain arbitrary
(i.e., we kept a \texttt{JUMP} between these observations' sets of
TOAs). Using the \texttt{DD} model \citep{DamourDeruelle1985,
  DamourDeruelle1986}, we find
$e=0.968$. ``Faking'' phase connection by removing the \texttt{JUMP}
yields $e=0.950$; alternatively, including arbitrary phase
\texttt{JUMP}s between all TOA sets yields $e \approx 0.8$, which we
take to be the lowest possible $e$ of this system.

\begin{figure}
\begin{center}
\includegraphics[scale=0.5]{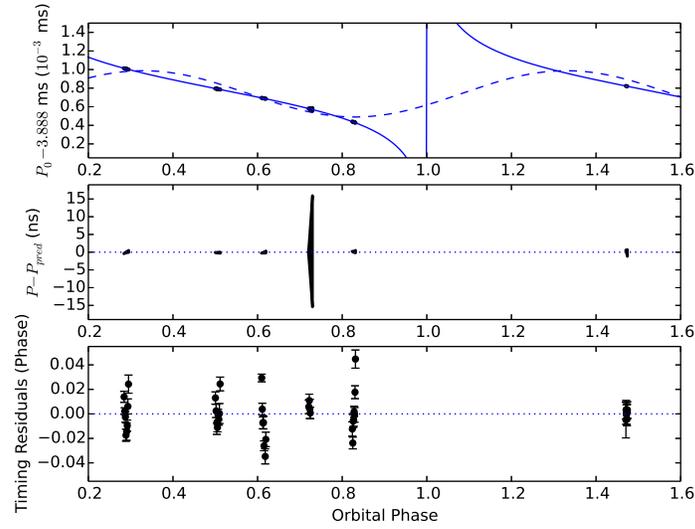}
\caption{\textit{Top:} Predicted spin periods for a circular orbit
  (dashed line) and the \texttt{DD} orbital model with a \texttt{JUMP}
  between observations 5 and 6 ($e = 0.967$; solid line).  The
  incoherently measured $P$ and $\dot{P}$, including uncertainties,
  are overlaid in black.  \textit{Middle:} Residuals from subtracting
  $P$ predicted by the $e = 0.967$ model from the measured $P$.
  \textit{Bottom:} Timing residuals from TOAs fit with the $e = 0.967$
  model.  We attribute the systematics in the TOAs to phase
  uncertainties which will be resolved with a fully phase-connected
  timing solution.}
\label{fig:residuals}
\end{center}
\end{figure}

The best-fit \texttt{DD} timing model parameters are in
Table~\ref{tab:timing}, with fit residuals in
Figure~\ref{fig:residuals}.  The systematics in the residuals
  are present in all our fits, including those with \texttt{JUMP}s
  between all observations; we attribute them to parameter, and
  therefore phase, uncertainties resulting from the lack of TOAs
  through periastron.  We stress that the initial timing parameters
in Table~\ref{tab:timing} belong to a family of solutions---a unique
determination of the MSP's timing solution requires further
observations, especially through periastron passage. If further timing
confirms the parameters, then NGC6652A will be the most eccentric
binary MSP known to date.

\section{Discussion \label{sec:discussion}}
We adopt timing parameters from the $e = 0.950$ model, pulsar mass
$m_{\mathrm p} = 1.4\,M_{\odot}$, and cluster parameters from
\citet[][2010 edition]{Harris1996} for all calculations, unless
otherwise stated.

\subsection{Cluster Membership \label{sec:membership}}
The discrepancy between the discovery and predicted DMs (63.35 and
190\,cm$^{-3}$\,pc, respectively) initially led us to question the
MSP's cluster association \citep{DeCesarRR2011}.  However, the Cordes
\& Lazio model commonly has uncertainties of a factor 0.5--2, and
sometimes larger.  The measured DM is consistent with the low optical
reddening $E_{B-V} = 0.10 \pm 0.02$ \citep{OrtolaniBB1994} and
estimated X-ray absorption column $N_{\mathrm H} \sim$\,$5.5 \times
10^{20}$\,cm$^{-2}$ \citep{PredehlSchmitt1995} toward NGC\,6652.

The high $e$ of NGC6652A is much more probable in a GC than in the
field due to the high probability of stellar encounters
\citep{CamiloRasio2005}, discussed further below.  Additionally, given
the beamwidth of the GBT at S-band ($6.3^{\prime}$), we estimate a
$\approx$\,0.2\% chance of finding an unassociated MSP coincident with
NGC\,6652 (assuming an isotropic distribution of known galactic MSPs).
We conclude that the MSP is almost certainly a cluster member.

\subsection{Nature of the Companion \label{sec:nature}}
The minimum companion mass (orbital inclination $i = 90^{\circ}$) is
$m_{\mathrm{c, min}} \simeq 0.74\,M_{\odot}$ (Table~\ref{tab:timing}).
Comparing with the Australia Telescope National Facility (ATNF) Pulsar
Catalog\footnote{\url{http://www.atnf.csiro.au/people/pulsar/psrcat/}}
shows that the companion is unusually massive; it may be a main
sequence (MS) or evolved star, or a compact object.  Based on the
cluster's age ($11.7 \pm 1.6$\,Gyr; \citealt{ChaboyerSA2000}), the
main-sequence turn-off mass is $\approx$\,$0.8\,M_\odot$
\citep{Stacey+2012}.  For $i < 70^{\circ}$, $m_{\mathrm c} >
0.8\,M_\odot$, limiting the range of inclinations for which an
unevolved MS companion is possible \citep[c.f.][]{FreireRG2007}.

For a non-compact companion, significant tides at periastron will
circularize the orbit.  The circularization, or dissipation, timescale
$t_{\mathrm D}$ for an eccentric binary system can be estimated as
\citep[][and references therein]{SocratesKD2012}
\begin{equation}
t_{\mathrm D} \equiv \frac{m_{\mathrm c} a^8_{\mathrm F}}{3 k_{\mathrm L} \tau G m^2_{\mathrm{p}} R^5}
\end{equation}

\noindent where $a_{\mathrm F} \equiv a \sin{i}\,(1-e^2)$, $\tau$ is
the constant tidal lag time of the companion, $k_{\mathrm L}$ is the
Love number, and $R$ is the companion's radius.  For high $e$, the
tidal quality factor $Q$ is related to $\tau$ by Equation~23 of
\citet{SocratesKD2012}.  We estimate $k_{\mathrm L}$ to be between
0.05--0.15 for both MS and WD companions, based on calculations with
Modules for Experiments in Stellar Astrophysics
\citep[\textsc{mesa};][]{Paxton+2011, BrookerOlle1955}.  For $Q =
10^6$, the circularization timescales are $\sim$\,Myr for a MS
companion and $\sim$\,$10^4$\,Gyr for a WD companion.  We conclude
that the companion is a compact object, whose mass and nature will be
constrained through further timing.

The merger timescale from gravitational wave-driven inspiral
  depends on $e$ as $t_{\mathrm{merge}}\propto(1-e^2)^{7/2}$
  \citep{Peters1964}.  For $i=90$
  ($m_{\mathrm{c,min}}=0.74\,M_{\odot}$) and $e=0.95$,
  $t_{\mathrm{merge}}\approx12$\,Gyr; varying $e$ yields a range of
  $t_{\mathrm{merge}}$\,$\sim$\,1\,Gyr ($e=0.975$) to $>100$\,Gyr
  ($e\la0.9$).  The system may therefore be disrupted
(Section~\ref{sec:exchange}) before it has time to merge.  In the
event of a merger, when the system comes into contact, the outcome
will depend on the exact nature of the binary.  Stable mass transfer
will be possible for $q \equiv m_{\mathrm c}/m_{\mathrm p} < 2/3$ ($i
> 50^{\circ}$), forming an ultra-compact X-ray binary and possibly an
isolated MSP.  For larger inclinations, the mass transfer will
  be unstable; while a black hole would form from accretion-induced
  collapse \citep[AIC;][and references therein]{GiacomazzoPerna2012}
  if the system mass exceeds the maximum NS mass, it is unclear
  whether substantial mass would be ejected from the system during
  unstable mass transfer, preventing AIC (L. Bildsten, \textit{private
    comm.}).  An eventual merger of this system may result in a
  long-GRB-like, calcium-rich transient \citep[e.g.,][]{KingOD2007} if
  the companion is a massive WD, or a short GRB for a NS companion
  \citep[e.g.,][]{GrindlayZM2006}.

\subsection{Post-Keplerian Parameters and Mass Constraints \label{sec:mass}}
Finding $m_{\mathrm c}$ and $m_{\mathrm p}$ requires measurements of at
least two post-Keplerian (PK) parameters \citep[we employ the general
  relativistic formalism of][]{DamourTaylor1992}.  Our preliminary
timing solution predicts that the rate of change of the longitude of
periastron passage $\omega$ (i.e., the orbital precession rate) is
$\dot{\omega}$\,\,$ > 0.08\degr$\,yr$^{-1}$, the Einstein delay
$\gamma_{\mathrm{GR}} > 10$\,ms, and $\dot{P}_b > 6 \times
10^{-12}$\,s\,s$^{-1}$.  Because of the high $e$, we will measure
$\dot{\omega}$ with high precision: from simulations assuming
  the $e=0.95$ orbital model parameters, we find that we will measure
$\dot{\omega}$ with $>100\sigma$ significance after one year of
timing, yielding the total system mass ($\dot{\omega} \propto
M_{\mathrm{tot}}^{2/3}$, where $M_{\mathrm{tot}} = m_{\mathrm p} +
m_{\mathrm c}$) and constraints on $m_{\mathrm p}$ and $m_{\mathrm
  c}$. Knowledge of the pulsar position (requiring 1\,yr of
  timing or an interferometric detection) would yield a measurement of
  $\dot{\omega}$ with one month of phase-connected timing. Our
  simulations also show that $\gamma_{\mathrm{GR}}$ will be measured
  with 10\% uncertainty with 2.5\,yr of phase connection.

We may also measure one Shapiro delay parameter, $s = \sin{i}$.  For
$i > 50^{\circ}$, the timing residuals from $s$ are significantly
larger than the $\approx$\,$20$\,$\mu$s uncertainties in the pulse TOAs
we used to build the timing model.  Statistically, it is most likely
that the MSP companion is a WD, requiring $i > 40\degr$ for
$m_{\mathrm c} < 1.4$\,M$_{\odot}$.  Even a marginal detection of
Shapiro delay will yield a precise $s$ because $\dot{\omega}$ and $s$
are nearly orthogonal in the mass-mass diagram
\citep{Lynch+2012}.  With these two PK measurements, we would
precisely measure $m_{\mathrm p}$ and $m_{\mathrm{c}}$.  We note that
the very precise mass of PSR J1807$-$2500B (NGC6544B) was measured in
this way \citep{Lynch+2012}.

\subsection{System Origin \label{sec:origin}}
The vast majority of field MSP binaries have circular orbits
(cf.\ \citealt{Champion+2008}) from dissipation during the
mass-transfer phase \citep{Phinney1992}; known eccentric systems in
the field are either double NSs (with eccentricity coming from a
second supernova (SN) kick; e.g., \citealt{BrandtPodsiadlowski1995}),
disrupted triples \citep{Champion+2008}, or possibly NS-He WD
binaries with circumbinary disks \citep[][and references
  therein]{Antoniadis2014}.  In contrast, a number of the MSP binaries
in GCs are substantially eccentric \citep[e.g.,][]{Freire+2008b,
  Lynch+2012}\footnote{Also see
  \url{http://www.naic.edu/{\til}pfreire/GCpsr.html}.}, with a likely
origin in dynamical encounters (e.g., \citealt{VerbuntFreire2014}).
The highest-$e$ binary MSP currently known, PSR\,J0514$-$4002A
\citep[NGC1851A;][]{FreireRG2007}, has $e = 0.888$ and an unusually
massive ($m_{\mathrm c} > 0.96\,M_{\odot}$) companion, similar to
NGC6652A. Here we consider the plausibility of several mechanisms
through which NGC6652A could have gained its high $e$.

\subsubsection{Possible Formation Mechanisms}
There are several ways to form a high-eccentricity system like
NGC6652A.  An initially circular orbit may gain eccentricity from
3-body encounters with other stars in the GC \citep{RasioHeggie1995}.
For a double neutron star (DNS), the eccentricity could have been
imparted on the system by the SN kick of a massive companion
\citep[][and references therein]{BrandtPodsiadlowski1995}.  The system
could also have formed through an exchange encounter, in which the
original companion was ejected from the system and the third body
became the new companion \citep[e.g.,][]{VerbuntFreire2014}.  In this
case, the new companion can be any type of compact object.

The SN kick is ruled out by observational evidence that all known
radio pulsars with NS companions have spin periods of 20--100\,ms
\citep{Tauris2011}, suggesting that MSPs cannot be fully recycled by
short-lived, massive companions.  The first scenario is plausible, as
using Equation 5 of \citet{RasioHeggie1995}, we find that $\approx
11.8$\,Gyr (comparable to the GC age) of non-exchange three-body
interactions would be needed for a binary in an initially circular
orbit to gain $e = 0.95$. However, the exchange encounter scenario
seems most natural, and we discuss this mechanism in more detail
below.

Other scenarios for the origin of the binary's eccentricity include a
physical collision between a MSP and a giant star
\citep[e.g.,][]{FreireRG2007}, or a triple system in which the
outermost companion is pumping the eccentricity of the inner binary
\citep[e.g., B1620$-$26;][]{Thorsett+1999}.  These mechanisms cannot
be excluded \textit{a priori}, but are outside the scope of this
letter.

\subsubsection{Dynamical Formation Through an Exchange Encounter \label{sec:exchange}}

We consider a dynamical encounter resulting in a companion exchange,
using \textsc{fewbody} \citep{Fregeau+2004} to simulate a particular
scenario.  As a progenitor system, we take the current most common MSP
binary in GCs: a MSP in a circular 2\,d orbit with a low-mass
companion. We chose $m_{\mathrm c} = 0.2\,M_\odot$, which follows from
a binary period of 2\,d using the period-core mass relation from
\citet{TaurisSavonije1999} for Pop~II stars.  We simulated 5000
encounters between this binary and a third body, drawing the incoming
velocities from a Maxwellian distribution\footnote{While the velocity
  profile of NGC~6652 has not been measured directly, we estimate a
  velocity dispersion $\sigma$ of about 10--15\,${\rm km\,s}^{-1}$,
  scaling from globular clusters with similar physical core radii
  (NGC~6388, 6093, and 6441); comparable values are obtained by
  \citet{McLaughlinvanderMarel2005}.} using $\sigma=10\,{\rm
  km\,s}^{-1}$ distributed between 0 and $30a$ (where $a$ is the
binary's semi-major axis).  For the third body we assume a WD with
$0.7\,M_\odot$.

Approximately 70\% of the encounters result in an exchange, with the
low-mass companion ejected and an eccentric binary remaining.  The new
binary has a range of eccentricies strongly biased toward high values,
with 66\% of the new systems having $e > 0.8$, but energies comparable
to that of the progenitor.  The orbit has expanded due to the factor
of 3.1 increase in $m_{\mathrm c}$, leading to a factor of
$3.1^{3/2}$ increase in $P_{\mathrm b}$.  Therefore, systems
with $e \approx 1$ and $P_{\mathrm b} \approx 10$\,d are naturally
formed through this mechanism.  
If no exchange happened, then the binary remains close to the 2\,d
initial period, albeit with enhanced eccentricity.

We estimate the frequency of encounters between a particular
NGC6652A-like binary and a single star in NGC\,6652 using the
single-binary encounter rate $\gamma$ from \citet{VerbuntFreire2014},
normalized to M4, and find $\gamma_{6652} \approx
6.7\gamma_{\mathrm{M4}}$.  A NGC6652A-like binary in M4 would
encounter single stars at a rate $\xi_{1+2} \sim
(\rho_{\mathrm{c,M4}}/L_{\odot}) \sigma_{1+2} v_{\mathrm{M4}}$, where
$\rho_{\mathrm c}$ is the GC core density and $\sigma_{1+2}$ is the
gravitationally-focused single-binary cross-section \citep[Equation
  A2,][]{LeighSills2011}.  In M4, this encounter rate is $\xi_{1+2,M4}
\sim 0.17$\,Gyr$^{-1}$; in NGC\,6652, $\xi_{1+2,6652} \sim
1$\,Gyr$^{-1}$. We note that the core radius of $1^{\prime\prime}.15$
measured by \citet{NoyolaGebhardt2006} is much smaller than that from
\citet[][2010 edition]{Harris1996}, yielding $\gamma_{6652}
\approx 38\gamma_{\mathrm{M4}}$ and $\xi_{1+2,6652}
\sim$\,6\,Gyr$^{-1}$.  The companion exchange scenario is therefore
quite plausible.  The position of NGC6652A in the cluster may
  give additional clues to its formation
  \citep[cf.][]{PhinneySigurdsson1991}.

\section{Conclusions \label{conclusions}}
We discovered one new MSP, NGC6652A.  Although NGC\,6388 and NGC\,6652
are expected to host substantial MSP populations, cluster MSPs are
extremely faint---detecting them requires long integration times and
the largest telescopes in the world.  We did not find more MSPs in
these GCs simply because we are sensitivity limited.  NGC6652A
  is an intriguing source for southern-hemisphere Square Kilometer
  Array (SKA) precursors and eventually the SKA Mid-Frequency Aperture
  Array\footnote{\url{https://www.skatelescope.org/mfaa}}.

Our timing analysis over 1.2\,orbit of NGC6652A shows that the
  MSP is in an extremely eccentric binary system with an unusually
  massive compact companion. The system quite plausibly formed
through an exchange encounter in the dense cluster environment.  We
cannot exclude all other formation mechanisms \textit{a priori};
determining the nature of the companion will help discriminate between
scenarios.  Similarly, a precise position will help determine whether
the binary is dynamically relaxed (and hence close to the core as
expected from mass segregation) or has been kicked out of the core by
a recent encounter.

With a phase-connected timing solution over $\geq 1$\,yr, we
  will uniquely determine the MSP's timing parameters and measure its
  position and $\dot{\omega}$ to high precision.  After
  $\approx2.5$\,yr of timing, we expect to measure
  $\gamma_{\mathrm{GR}}$, allowing measurements of $m_{\mathrm p}$ and
  $m_{\mathrm c}$ and clarifying the companion's nature.  If $i >
  50^{\circ}$, we may also measure $\sin{i}$.  New timing observations
  are underway, and will be reported upon in a subsequent paper.

\begin{acknowledgments}

The authors thank P. Arras, P. Freire, J. Fuller, V. Kalogera, and
F. Rasio for helpful discussions; F. Camilo for helpful discussions
and the use of his computer cluster; and the anonymous referee for
useful suggestions that improved the quality of this letter.  MED
acknowledges funding from NSF Award No.\ AST-1312822 and NASA's CRESST
grant No. 01526268. PSR is supported by the Chief of Naval
  Research (CNR). AMG is funded by a NSF Astronomy and Astrophysics
Postdoctoral Fellowship under Award No.\ AST-1302765. 


\end{acknowledgments}


\newpage



\begin{deluxetable}{ccccccc}
\tablewidth{0pt}
\tabletypesize{\footnotesize}
\small
\tablecaption{Targeted Globular Clusters\tablenotemark{$a$}}
\tablecolumns{7}
\tablehead{
\colhead{Cluster Name} & \colhead{$\alpha$} & \colhead{$\delta$} & \colhead{$l$} & \colhead{$b$} & \colhead{Distance} & \colhead{Predicted DM} \\
\colhead{} & \colhead{} & \colhead{} & \colhead{(degrees)} & \colhead{(degrees)} & \colhead{(pc)} & \colhead{(cm$^{-3}$\,pc)}
}
\startdata
NGC\,6388 & $17^{\mathrm h}36^{\mathrm m} 17\fs88$ & $-44\degr44^{\prime} 0\farcs24$ & $345.56$ & $-6.74$ & $11.6\pm2.0$\tablenotemark{$b$} & 340\tablenotemark{$c$} \\
NGC\,6652 & $18^{\mathrm h} 35^{\mathrm m} 44\fs86$ & $-32\degr 59^{\prime} 25\farcs10$ & $1.53$ & $-11.37$ & $9\pm1$\tablenotemark{$d$} & 190\tablenotemark{$c$}
\enddata
\tablenotetext{$a$}{The cluster positions were set to the optically determined positions of the cluster centers.}
\tablenotetext{$b$}{\citet{Moretti+2009}}
\tablenotetext{$c$}{\citet{CordesLazio2002}}
\tablenotetext{$d$}{\citet{ChaboyerSA2000}}
\label{tab:targets}
\end{deluxetable}

\begin{deluxetable}{lrrrrcrrrrc}
\tablewidth{0pt}
\tabletypesize{\footnotesize}
\small
\tablecaption{Observation Log\tablenotemark{$a$}}
\tablecolumns{11}
\tablehead{
\colhead{Date} &  \colhead{$t_{\mathrm{int}}$}  & \colhead{$S_{\mathrm{min}}$\tablenotemark{$b$}} & \colhead{S/N} & \colhead{$S_2$\tablenotemark{$c$}} & \colhead{$N_{\mathrm{TOA}}$} & \colhead{$t_{\mathrm{TOA}}$} & \colhead{$\langle \sigma_{\mathrm{TOA}} \rangle$} & \colhead{Barycentric $P$} & \colhead{Barycentric $\dot{P}$} & \colhead{$z$} \\
\colhead{}   & \colhead{(s)} & \colhead{($\mu$Jy)} & \colhead{} & \colhead{($\mu$Jy)} & \colhead{} & \colhead{(s)} & \colhead{($\mu$s)} & \colhead{(ms)} & \colhead{($10^{-12}$\,s\,s$^{-1}$)} & \colhead{(Fourier bins)}
}
\startdata
\sidehead{NGC\,6652}
2010 Oct 19 & 9470  & 5.8 & 25.2 & 22.5 & 8 & 1200 & 20 & 3.88937447(3) & $-1.223 \pm 0.027$ & 8.50  \\
2010 Oct 21 & 10062  & 5.6 & 14.1 & 12.2 & 8 & 1250 & 22 & 3.88915225(3) & $-0.888 \pm 0.023$ & 7.00 \\
2010 Oct 22 & 8878  & 6.0 & 5.8 & 5.4 & 7 & 1100 & 17 & 3.88904917(3) & $-0.916 \pm 0.028$ & 6.00 \\
2010 Oct 23\tablenotemark{$d$} & 7504  & 6.5 & 17.4 & 17 & 4 & 940 & 23 & 3.888933(1) & $-1.33 \pm 2.09$ & $\cdots$\tablenotemark{$d$} \\
2010 Oct 24 & 6701  & 6.8 & 14.2 & 15.1 & 8 & 840 & 21 & 3.88878963(5) & $-1.883 \pm 0.060$ & 6.25 \\
2010 Oct 29 & 3086  & 10.1 & 12.6 & 19.7 & 7 & 380 & 32 & 3.8891541(1)   & $-0.934 \pm 0.286$ & 0.50 \\[1ex]
\sidehead{NGC\,6388}
2010 Oct 21 & 6278 & 7.3 & $\cdots$ & $\cdots$ & $\cdots$ & $\cdots$ & $\cdots$ & $\cdots$ & $\cdots$ & $\cdots$ \\
2010 Oct 24 & 3403 & 9.9 & $\cdots$ & $\cdots$ & $\cdots$ & $\cdots$ & $\cdots$ & $\cdots$ & $\cdots$ & $\cdots$ \\
2010 Oct 29 & 4080 & 9.0 & $\cdots$ & $\cdots$ & $\cdots$ & $\cdots$ & $\cdots$ & $\cdots$ & $\cdots$ & $\cdots$ \\
2011 Feb 05 & 5412 & 7.8 & $\cdots$ & $\cdots$ & $\cdots$ & $\cdots$ & $\cdots$ & $\cdots$ & $\cdots$ & $\cdots$ \\
2011 Apr 08 & 6130 & 7.4 & $\cdots$ & $\cdots$ & $\cdots$ & $\cdots$ & $\cdots$ & $\cdots$ & $\cdots$ & $\cdots$ \\
2011 May 06 & 4338 & 8.7 & $\cdots$ & $\cdots$ & $\cdots$ & $\cdots$ & $\cdots$ & $\cdots$ & $\cdots$ & $\cdots$
\enddata
\tablenotetext{$a$}{All observations were taken at 2\,GHz with
  $\sim$\,700\,MHz effective bandwidth and 40.96\,$\mu$s time
  resolution.  The beamsize was $\sim$\,6\am.}
\tablenotetext{$b$}{$S_{\mathrm{min}}$ was calculated with the
  radiometer equation for pulsed signals \citep[Appendix A1.4
    of][]{LorimerKramer2005}, using $\mathrm{S/N_{min}} = 5$ and pulse
  width $W = 0.1P$.  For observations at S band, the correction factor
  $\beta = 1.05$, gain $G = 1.9$\,K\,J$^{-1}$, and $T_{\mathrm{rec}}
  \simeq 22$\,K; for these observations, the number of polarizations
  $n_{\mathrm p} = 2$, and effective bandwidth $\Delta f = 700$\,MHz.
  The sky temperature $T_{\mathrm{sky}} = 0.8$\,K for NGC\,6652 and
  1.4\,K for NGC\,6388 \citep{deOliveiraCosta+2008,
    deOliveiraCosta+2010}, giving $T_{\mathrm{sys}} = T_{\mathrm{rec}}
  + T_{\mathrm{sky}} = 22.8$\,K and $23.4$\,K, respectively.}
\tablenotetext{$c$}{The 2\,GHz flux density $S_2$ was calculated with
  the same radiometer equation parameters as for $S_{\mathrm{min}}$,
  but using the measured S/N rather than S/N$_{\mathrm{min}}=5$ and
  the measured $W = 0.0625P$ rather than $W = 0.1P$. The smaller
  measured $W$ explains why we find $S_2 < S_{\mathrm{min}}$ on 2010
  October 22.}
\tablenotetext{$d$}{The DM used for coherent dedispersion of individual channels was accidentally set to 9.0\,cm$^{-3}$\,pc for this observation. As a result, the MSP was not found in a straightforward acceleration search.}
\label{tab:observations}
\end{deluxetable}

\begin{deluxetable}{lcrcr}
\tablewidth{0pt}
\tabletypesize{\small}
\small
\tablecaption{NGC6652A Timing Solution\tablenotemark{$a$} \label{tab:NGC6652A_timing}}
\tablecolumns{5}
\tablehead{
\colhead{Timing Parameter} & & \colhead{With \texttt{JUMP}} & & \colhead{Without \texttt{JUMP}}
}
\startdata
Right Ascension\tablenotemark{$b$} (J2000.0) & & 18\h\,35\m\,44\s.856 & & 18\h\,35\m\,44\s.856 \\ [0.5ex]
Declination\tablenotemark{$b$} (J2000.0) & & $-32$\dg\,59\am\,25\as.08 & & $-32$\dg\,59\am\,25\as.08 \\ [0.5ex]
Dispersion Measure\tablenotemark{c} (cm$^{-3}$\,pc) & & 63.35 & & 63.35 \\ [0.5ex]
Spin period, $P$ (ms) & & 3.888824(1) & & 3.8888289774(4) \\ [0.5ex]
Spin period epoch (MJD) & & 55488.931354 & & 55488.931354 \\ [0.5ex]
Spindown rate\tablenotemark{$d$}, $\dot{P}$ (s\,s$^{-1}$) & & 0 & & 0 \\ [0.5ex]
Orbital period, $P_{\mathrm b}$ (days) & & 9.2460(5) & & 9.2459(5) \\ [0.5ex]
Projected semimajor axis, $x$ (s) & & 19.6(3) & & 19.09(5) \\ [0.5ex]
Eccentricity, $e$ & & 0.968(5) & & 0.950(1) \\ [0.5ex]
Epoch of periastron passage, $T_0$ (MJD) & & 55477.061(5) & & 55477.0401(6) \\ [0.5ex]
Longitude of periastron, $\omega$ (degrees) & & 291(1) & & 289.2(2) \\ [0.5ex]
Minimum companion mass, $m_{\mathrm{c,min}}$ ($M_{\odot}$) & & 0.765(14) & & 0.736(3) \\ [0.5ex]
Fit $\chi^2$ per degrees of freedom & & 347.9/35 & & 369.37/36
\enddata
\tablenotetext{$a$}{The solution uses the \texttt{DD} model
  \citep{DamourDeruelle1985, DamourDeruelle1986} and the TDB time
  system. The error on the last digit(s) of each parameter value is
  denoted in parentheses. The true solution is one in a family of
  solutions represented by the parameters listed here. The middle
  column gives the timing parameters obtained when allowing an
  arbitrary number of pulsar rotations between the fifth and sixth
  observations.  The right column gives the parameters obtained with
  ``forced'' phase connection.  The unique solution will be determined
  with further timing observations that include a periastron passage.}
\tablenotetext{$b$}{The position was fixed at the cluster's center.}
\tablenotetext{$c$}{The DM was fixed to this best value from the
  discovery observation (2010 October 19).}
\tablenotetext{$d$}{The spindown rate was fixed at zero; a
  phase-connected timing solution spanning $\sim$\,one year will
  measure this parameter.}
\label{tab:timing}
\end{deluxetable}


\end{document}